\documentclass[prl,aps,twocolumn]{revtex4}

\usepackage{graphicx}

\begin{document}

\title{Quasioptical qualification of parallel/series arrays of cold-electron bolometers with a cross-slot antenna}
\author{A.S. Mukhin$^{1}$}
\author{A.V. Gordeeva$^{1,2}$}
\email{anna.gord@list.ru}
\author{L.S. Revin$^{1,2,3}$}
\author{A.E. Abashin$^{1}$}
\author{A.A. Shishov$^{1}$}
\author{A.L. Pankratov$^{1,2,3}$}
\author{S. Mahashabde$^{4}$}
\author{L.S. Kuzmin$^{1,4}$}

\affiliation{$^1$Nizhny Novgorod State Technical University n.a. R.E. Alekseev, GSP-41, Nizhny Novgorod, 603950, Russia \\
$^{2}$Lobachevsky State University of Nizhny Novgorod,
GSP-20, Nizhny Novgorod, 603950, Russia\\
$^{3}$Institute for Physics of Microstructures of RAS,
GSP-105, Nizhny Novgorod, 603950, Russia\\
$^{4}$Chalmers University of Technology, 41296, Gothenburg, Sweden}

\begin{abstract}
We perform experimental and theoretical study of the parallel-series arrays of Cold-Electron Bolometers (CEBs) integrated into a cross-slot antenna and composed with an immersion silicon lens. The purpose is to determine the absorption efficiency, the responsivity and the noise equivalent power (NEP) of the bolometers. The absorbed power has been found in two independent ways. The comparison of two approaches gives better understanding of the system and secures from misinterpretations. The first approach is fitting of the bolometer IV curves with solutions of heat-balance equations. The second approach is modeling of electromagnetic properties of the system, including an antenna, lens, optical can, band-pass filters and black body source. The difference between both methods does not exceed $30\%$. The optimization of experimental setup is proposed to approach the photon limited detection mode.

\end{abstract}



\maketitle

\section{Introduction}

The cold-electron bolometers (CEBs) \cite{Kuzmin2002, Kuzmin2012} are promising candidates as detector systems for various space and balloon missions due to their unique features, such as high sensitivity and broad dynamic range (due to electron cooling effect) . Another particular feature of CEB is insensitivity to Cosmic Rays (due to tiny volume of an absorber and strong decoupling of phonon and electron subsystems) \cite{Sala}.  Importance of the immunity of CEBs to CR was especially understood after serious problems of Planck with glitches due to underestimated influence  of cosmic rays through the substrate \cite{Planck}.

For operation under high optical power load, the parallel/series arrays of CEBs have been proposed to keep high sensitivity by distributing power between bolometers of the array \cite{Kuzmin2008}. In the present paper we perform extensive analysis of the parallel/series arrays of CEBs developed for 350 GHz channel of the BOOMERanG balloon telescope \cite{Masi2006}. The arrays were
combined with the cross-slot antenna \cite{Zmuidzinas1998, Tarasov2011}. We have made measurements, electro-magnetic simulations and comparison with a theoretical heat-balance model \cite{Golubev2001}. The experiment has been done with room temperature JFET amplifiers.

The aim of the present paper is to find the optimal range of parameters for CEBs, which could show a minimum NEP for a given power load in correspondence to requirements of BOOMERanG. The input requirements for detectors are: 1) the optical power load $10$ pW for two polarizations or $P=5$ pW for one polarization (corresponding photon noise NEP$_{phot} = \sqrt{2 P hf} = 4.3\cdot 10^{-17}\textrm{W Hz}^{-1/2}$); 2) NEP of the detectors is less or equal to NEP$_{phot}$.
Using the idea of parallel/series arrays of bolometers \cite{Kuzmin2008} it has been found in the framework of the model \cite{Golubev2001}, that bolometers with the following parameters should meet the requirements of the mission: 6 bolometers in the array for one polarization channel, normal resistance of the array is $12$ kOhm, resistance of the absorber is $90$ Ohm.

The paper consists of four parts. First we discuss the electromagnetic model, performance of antenna in cases of fully matched and mismatched impedances and estimate absorption efficiency. In the second part we describe the experiment, fit results with the heat-balance model \cite{Golubev2001} and find in such a way the absorbed power.  In the third part we calculate the absorbed power in another way - using the information about electromagnetic properties of the system. Finally we calculate an electrical NEP and discuss what can be done to minimize it further to the level of the photon noise.

\section{Electromagnetic model}

\begin{figure}[h]
\resizebox{1\columnwidth}{!}{
\includegraphics{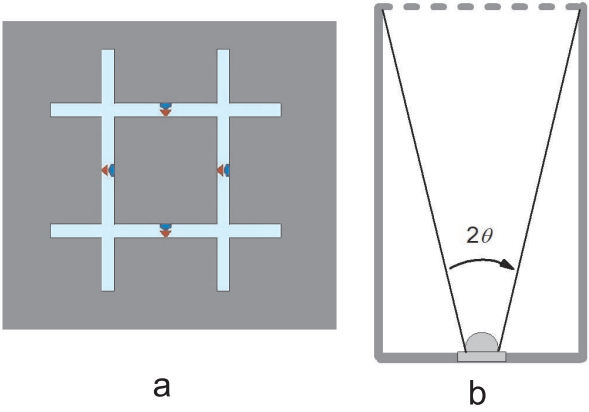}}
{\caption{a: Cross slot antenna with four uncorrelated ports. b: Scheme of a sample holder: the lens and the sample are placed inside the can with band-pass filters on top. }
\label{CS}}
\end{figure}

The presented design of cross-slot antenna is based on original work for 550 GHz \cite{Zmuidzinas1998}. For our purpose the antenna has been scaled down to 350 GHz (Fig.\ref{CS}). Each slot is coupled to an array of bolometers, absorbing the signal. The bolometer responses of opposite arrays are summed up for DC read out.

A lens coupled to the antenna significantly improves its performance. It has been shown in \cite{Filipovic1995} that the efficiency of a planar antenna with lens and antireflection (AR) coating can be as good as $70 \%$.

The antenna is made of gold with thickness 150 nm, shown as a gray color in schematic of Fig. \ref{CS}. Blue and red elements (real and imaginary parts of the impedance) in the middle of slots represent four arrays of bolometers connected in series for DC current and in parallel for AC current. In our setup the antenna works as a multiport antenna (two uncorrelated ports for each polarization).

The aim of electromagnetic simulations is to estimate how much power is absorbed in the bolometers. The closer we reconstruct the experimental environment the more accurate the result will be. In the experiment the sample is placed inside a metallic can with filters mounted on top. The bandwidth of the band-pass filters,  developed in Cardiff University, is 33 GHz. A schematic view of the sample holder is shown in Fig.\ref{CS}. The can is radiated by a black body source. The height of the metallic can is such that the antenna receives radiation from angles less that $2 \theta = 16^\circ$.

In simulations we send a plane wave with amplitude of electrical field $E_0$ at different angles $\theta$ and monitor current and voltage response in ports at 350 GHz. The distance between lens and the plane wave source is taken 9 mm in order to fit into the experimental angle, formed by the can. At this distance the area of the radiated plane is about 16 ${\rm mm}^2$.

\begin{figure}[h]
\resizebox{1\columnwidth}{!}{
\includegraphics{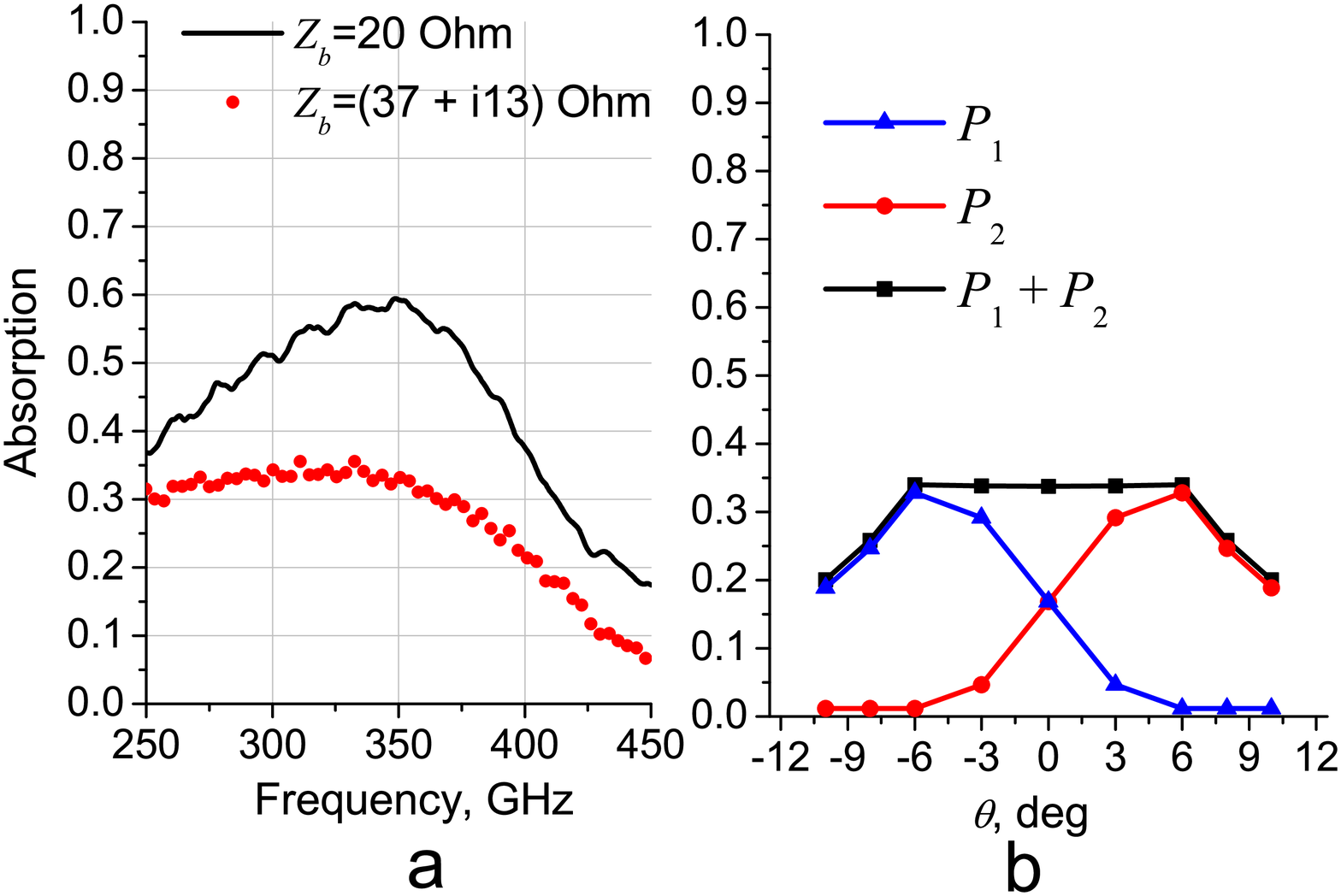}}
{\caption{ a: Absorbed power in two slots normalized to incident power versus frequency: black solid curve - fully matched impedance and AR coating, red dots - realistic impedance of bolometers and without AR coating. b: Absorbed power (mismatched, w/o AR) versus angle of incidence: P$_1$ and P$_2$ - in 1st and 2nd slots correspondingly, and the sum of absorption in two slots.}
\label{Can}}
\end{figure}

\begin{figure}[h]
\resizebox{1\columnwidth}{!}{
\includegraphics{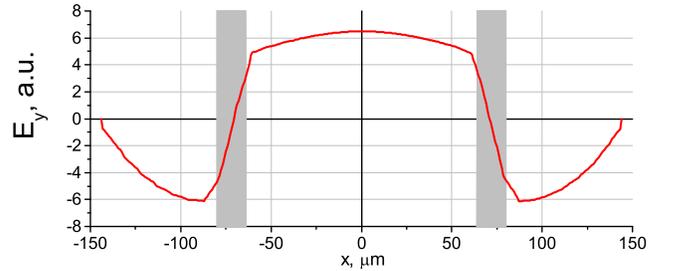}}
{\caption{Distribution of electric field in the slot at 350 GHz. Gray color - area inside slots.}
\label{Mode}}
\end{figure}

The absorption in two parallel slots is shown in Fig. \ref{Can}a as a function of frequency for normal incidence of the plane wave. This plot shows what maximum efficiency we can expect if we have full matching and AR coating. The resonance at 350 GHz (black solid curve) is clearly visible when the port impedance is matched with the slot impedance 20 Ohm. This resonance corresponds to EM field distribution shown in Fig. \ref{Mode}. The losses $40 \%$ are divided roughly half by half between the two factors: optical coupling of spherical lens with a plane wave and backside radiation.

The red dots in Fig. \ref{Can}b is the absorbtion under our experimental conditions. Additional $26 \%$ losses in this case are due to mismatched impedance ($10 \%$) and due to absence of AR coating ($16 \%$).

The real part of bolometer impedance $Z_b$ in one slot is 37 Ohm for the measured sample. Im[$Z_b$] is nonzero due to capacitive elements in series with resistance in CEBs, increasing the mismatch with the antenna impedance, which is purely resistive in the center of slots. The capacitance of aluminium junctions is reported to be from $1.6 \, \mu F/cm^2$ \cite{McCarthy1999} to $5 \, \mu F/cm^2$ \cite{Lichtenberger1989}. For our estimations we take the value $3\, \mu F/cm^2$. The area of the junctions is $A = 0.76 \, \mu m^2$. Taking into account that three bolometers are connected in parallel, the impedance of CEBs in one slot at frequency 350 GHz is $Z_b = 37 + i13$ Ohm.

The radiated power is equal to
\begin{equation}
P_{rad} = \langle S \rangle A = \frac{\varepsilon_0 c}{2} E_0^2 A,
\end{equation}
where $A$ is the area of radiated surface and $\langle S \rangle$ is a time-averaged Poynting vector.

The power dissipated in the port is $P_{abs} = |V| |I| \cos(\phi)/2$, where $\phi = \tan^{-1}(1/2\pi f R C)$ is a difference between voltage and current phases. Within the bandwidth the current and voltage amplitudes can be taken constant.
The powers absorbed in each slot with one polarization and its sum as a function of angle of incidence is shown in Fig. \ref{Can}b. In the case of normal incidence the maximal intensity is between the slots. Thus, the total absorbed power is nearly constant within angles $8^\circ$.

All simulations have been done for perfect alignment between lens and antenna. In experiment we cannot avoid some misalignment but we can guaranty that it is less than $50 \mu m$. If the lens center is shifted from the antenna center, one slot gets more power than another one. But the total response is an average between them so that the misalignment is not crucial for absorption in contrast to the polarization resolution, which is not considered here.

\section{Experiment and fitting}

In order to find the responsivity $S_V$ and the noise equivalent power we need to measure the bolometer response $dV$ to incoming radiation. Let us define responsivity and electrical NEP through the absorbed power $P_{abs}$:

\begin{equation}
S_V = dV/dP_{abs},\, \textrm{NEP} = \upsilon_n/S_V,
\end{equation}
where $\upsilon_n$ is the total measured voltage noise.

\begin{figure}[h]
\resizebox{1\columnwidth}{!}{
\includegraphics{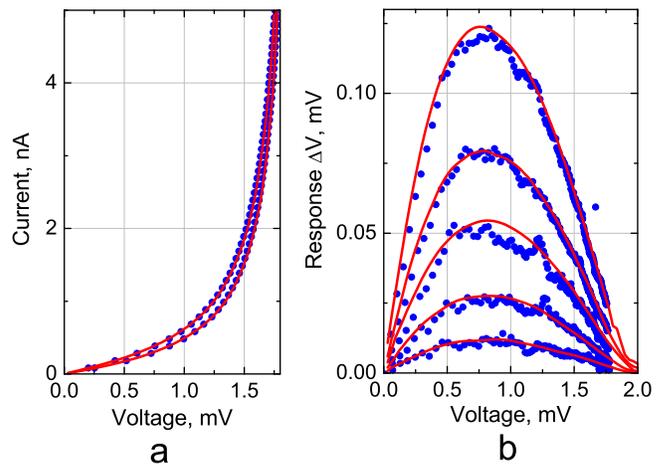}}
{\caption{a: IV curves at two temperatures of BB: 0.3 K and 7.1 K and b: response of bolometers at several temperatures of BB: 3, 3.8, 4.8, 5.9 and 7.1 K (from bottom to top). Blue markers - experiment, red lines - fit. Background power is 0.1 pW.}
\label{IVs}}
\end{figure}

The bolometers have been fabricated in Chalmers University of Technology. The sample has been cooled down to 300 mK and illuminated by a black body (BB) at temperatures from 0.3 K to 7K.  The typical response of bolometers is presented in Fig. \ref{IVs} for several temperatures of the black body: 3, 3.8, 4.8, 5.9 and 7.1 K.
The phonon temperature of the sample was monitored during measurements using on-chip SIN-array thermometers \cite{Kuzmin_thermo2007, Agulo_thermo2008} fabricated in the same vacuum technological cycle as bolometers. The shown response is purely optical response. For the temperatures of the black body above 7 K, we see mixed response to optical power and to additional phonon heating of the substrate due to increase of the base temperature above 300 mK.

The experimental IV-curves have been fitted using two heat balance equations (HBEs): for normal metal absorber and for superconducting electrodes. The model with one HBE for normal metal developed in Ref. \cite{Golubev2001} has been added with a $\beta$-term \cite{Mahashabde2015} and extended to two HBEs according to Ref. \cite{ONeil2011}. More details can be found in Ref. \cite{Mohamed2014}.

The model has many parameters, and not all of them can be measured directly. Therefore, fitting of experimental curves is a delicate task. However, a specific form of experimental current-voltage curves does not allow to vary fitting parameters in a broad range, and we believe that the most important parameters, such as the background and absorbed powers, are determined with sufficient accuracy. The normal resistance $R_N = 8$ kOhm and absorber resistance $R_{abs} = 110$ Ohm were taken from the experiment. The material constants $\Sigma_N = 1.25\,\, {\rm nW} {\rm m}^{-3}{\rm K}^{-5}$, $\Sigma_S = 0.3\,\, {\rm nW} {\rm m}^{-3}{\rm K}^{-5}$ and the critical temperature $T_c = 1.47$ K were found from the fit.

The response shape depends on a particular design and technological parameters like electrode thickness and  is determined mainly by interplay between three terms of HBEs:  power deposited in normal metal $P_n$, power deposited in superconductor $P_s$ and a fraction $\beta P_s$ which returns to the normal metal.

\begin{figure}[h]
\resizebox{1\columnwidth}{!}{
\includegraphics{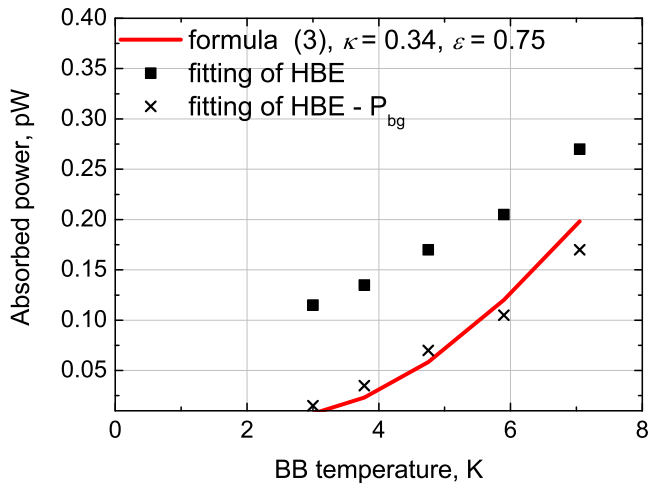}}
{\caption{ Absorbed power calculated in two different ways, see the discussion in the text.}
\label{Absorbed power}}
\end{figure}

The obtained fit qualitatively describes all main features of the IV-curves and gives very good quantitative agreement.
The power $P_{abs}$ found from the HBE model increases with the BB temperature as shown in Fig. \ref{Absorbed power} by square markers.
The important parameter is a background power $P_{bg}$, associated both with radiation from cryostat shields, as well as with various noise sources, effectively heating the bolometers. According to simulations, the bolometers absorbed $0.1$ pW before we started heating of BB. $P_{bg}$ is constant for all temperatures of BB and does not contribute to responsivity, but has to be included into the total system noise, since it leads to smoothing of the IV-curves.

The responsivity of CEBs monotonically decreases as a function of the absorbed power. The average responsivity equals $7.8\cdot 10^8$ V/W in the power range (0.1 - 0.27) pW in the most sensitive bias point $0.3$ nA.

The fitting using the heat balance equations gives a straightforward way to find the absorbed power, based on the results of measurements. An alternative approach will be considered in the next section. Comparison of two approaches is a test on validation of our assumptions and models.

\section{Absorbed power in EM model}

The efficiency of absorption $\kappa$, estimated in EM simulations for several angles of incidence, will be used below to calculate the absorbed power without heat balance equation, following works \cite{Kerr1997} and \cite{Tarasov2011}.
The coefficient $\kappa$ includes losses due to reflection from the lens and mismatch of impedances. Unaccounted losses are absorption in Si substrate and the lens.

The power radiated by a black body at temperature $T$ within a bandwidth $\delta f  \ll f$ into a single mode is:
\begin{equation}
P_{rad} = \varepsilon \frac{hf}{\exp(hf/kT)-1} \delta f, \label{Pabs1}
\end{equation}
where $\varepsilon$ is the emissivity coefficient of the black body. The bandwidth of our filters is $\delta f = 33$ GHz. Assuming the absorption is constant within the bandwidth, the absorbed power is $P_{abs} = \kappa P_{rad}$.

The advantage of this approach is that it is based on quasioptical properties of the system only and does not need microscopic parameters of bolometers. However, it does not account for a background power due to various noise sources. The absorbed power in absolute units is shown in Fig.\ref{Absorbed power} for both HBE and EM methods. Moreover, in order to compare them, the background power has been subtracted from the HBE result (crosses). A good agreement with HBE results is reached by variation of unknown emissivity in the range $0.7-0.9$. The red solid curve in Fig. \ref{Absorbed power} corresponds to the BB emissivity $\varepsilon = 0.75$, while the expected emissivity coefficient of cone copper BB covered with stycast should be around $0.9$. The possible explanation of discrepancy might be an inaccurate determination of the absorption efficiency and/or the BB temperature. The real temperature might be lower than we believe, since the thermometer is located at the copper leads rather than at the BB surface.

\section{Discussion}

\begin{figure}[h]
\resizebox{1\columnwidth}{!}{
\includegraphics{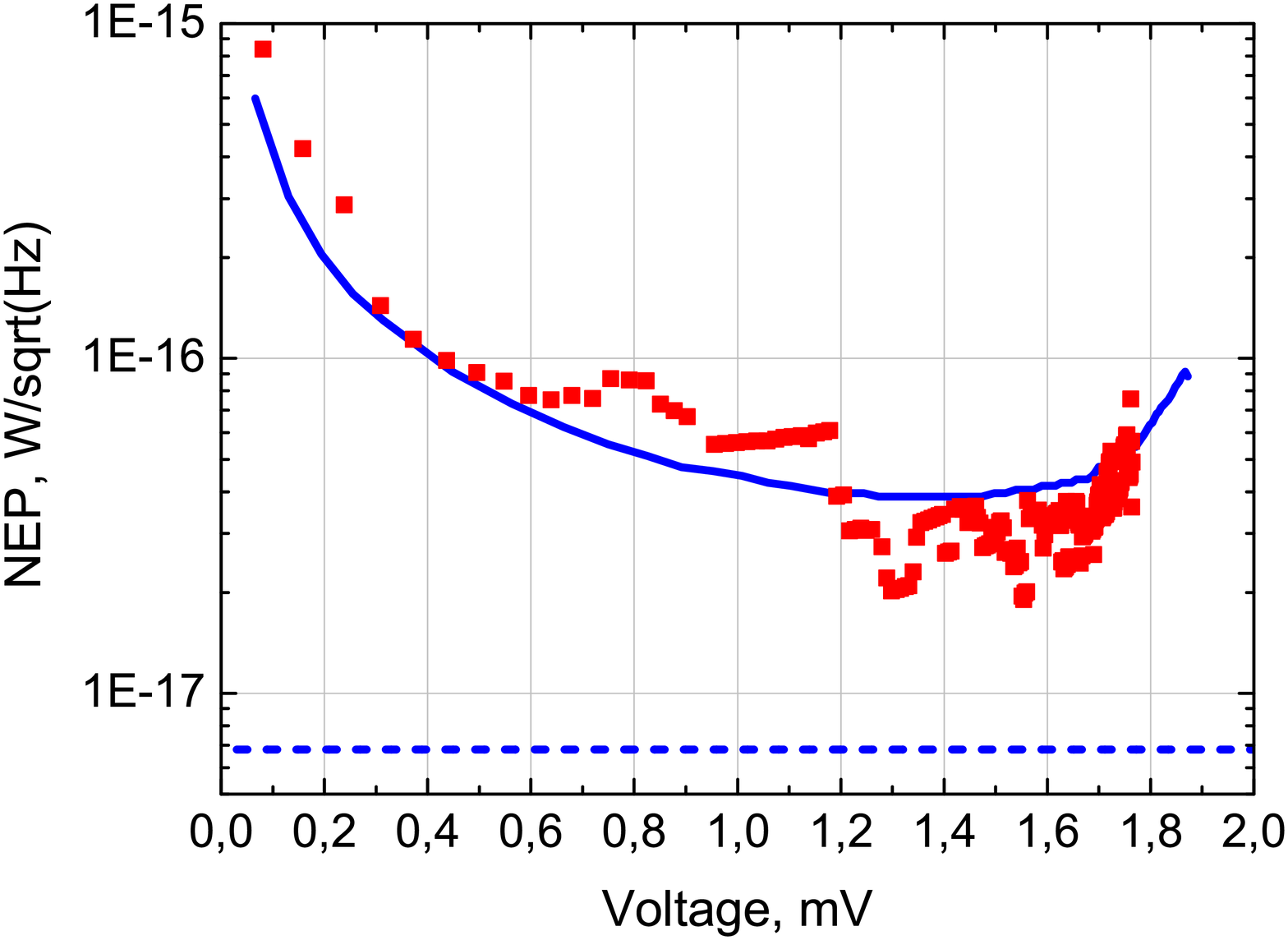}}
{\caption{ NEP measured (red markers) and calculated from the extended model \cite{Golubev2001,Mahashabde2015} (blue curve). Dashed blue line -- the level of photon noise for the calculated absorbed power.}
\label{NEP}}
\end{figure}

The electrical NEP of the detectors is calculated in Fig. \ref{NEP} for $P_{abs}=0.1$ pW: red points are based on the measured noise, blue curve is theoretical. The voltage noise for this plot has been measured at 120 Hz.  The dashed line is the photon noise level NEP$_{phot} = 7\cdot 10^{-18}\textrm{W Hz}^{-1/2}$. The minimal NEP is $4\cdot 10^{-17}\textrm{W Hz}^{-1/2}$ and the best ratio NEP/NEP$_{phot}$ is 6.

The parameters of bolometers predicted as optimal by the model \cite{Golubev2001} and actual measured ones are listed in table \ref{tab1}. The technological parameters $R_N$ and $R_{abs}$ are close to the requirements. Whereas both incident and absorbed powers are far away from the goal. What can be done to get us close to the goal NEP$_{tot}/$NEP$_{phot} \leq 1$?
\begin{table} [h]
  \centering
  \caption{List of parameters}\label{tab1}
  \begin{tabular} {l|c|r}
  \hline
  parameter & required & measured\\
  \hline
  $R_N$, kOhm & 12 & 8 \\
  $R_abs$, Ohm & 90 & 110 \\
  incident power, pW & 5 & 0.35 \\
  absorbed power, pW & 5 & 0.1 \\
  NEP$_{tot}/$NEP$_{phot}$ & 1 & 6 \\
  \hline
\end{tabular}
\end{table}

The detectors investigated here are designed for power load $5$ pW and equivalent NEP$_{phot} = 4.3\cdot 10^{-17}\textrm{W Hz}^{-1/2}$. In order to give $5$ pW of incident power our black body has to be heated up to 20 K.

\begin{figure}[h]
\resizebox{1\columnwidth}{!}{
\includegraphics{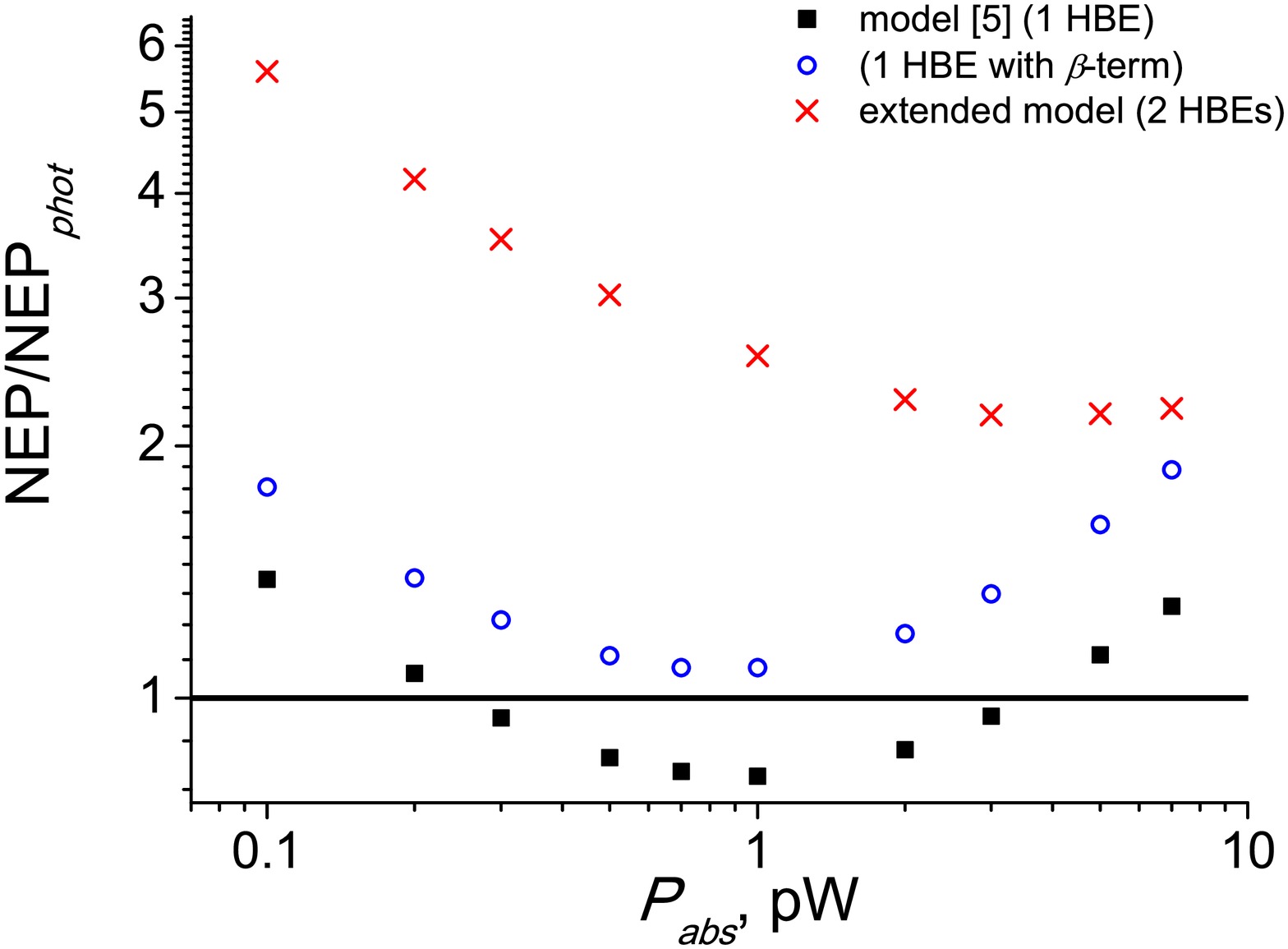}}
{\caption{ Predictions of NEP given by different models. Black line marks the level 1.}
\label{NEPratio}}
\end{figure}

In Fig. \ref{NEPratio} we show how the ratio between NEP and NEP$_{phot}$ changes with the absorbed power. We found that only the extended model with two HBEs describes our experiment well. Previous models give very optimistic but underestimated NEP according to the last experiment. Therefore, our future plans will be based on the extended model, which says that for powers more that 2 pW the photon noise is twice smaller than the total noise of the system. In this situation the photon noise can be still detected. The next experiments will be dedicated to this task.

We have discussed already that the model \cite{Golubev2001} needs to be extended to give more accurate bolometer characteristics. Besides, the efficiency of absorption is $0.34$ in our case instead of idealistic 1, which is not reachable even with an antireflection coating. Therefore, the bolometers have to be optimized for lower powers taking into account losses around $50 \%$. From the other hand the absorption efficiency has a potential to be improved as well due to better matching between antenna and bolometers.
Finally, the radiated power needs to be increased. New experiments with more efficient black body and better thermal decoupling are on the way. Larger absorbed powers will increase the photon noise level leading the desired NEP$_{tot}/$NEP$_{phot} \to 1$.

\section{Conclusion}

We have analyzed the performance of cold electron bolometers integrated with cross slot antenna at $300$ mK using two different approaches.
These methods have been compared, which allowed us to create the complete picture of the experiment. The difference between electromagnetic approach and heat balance equation approach is less than $30 \%$ if we account for the background power.

The modified HBE model \cite{Mahashabde2015,Mohamed2014} fits experimental curves perfectly. We have found that despite of numerous unknown parameters, there is only narrow range of parameters, giving good fits. We have shown that the electromagnetic method should be completed with calculation of the background power.

The design, considered here -- cross slot antenna with three bolometers in one slot, is close to the designs studied in other works \cite{Tarasov2011} (cross slot antenna with five bolometers in one slot) and \cite{TarasovBB} (twin-slot antenna with three bolometers). In \cite{Tarasov2011} the responsivity up to $8\cdot 10^8$ V/W and NEP as low as $2\cdot 10^{-17}\textrm{W Hz}^{-1/2}$ was reported in the temperature range 100-200 mK. In \cite{TarasovBB} $S = 3\cdot 10^8$ and NEP$=2\cdot 10^{-17}\textrm{W Hz}^{-1/2}$ were measured at $T=280$ mK. In our case the responsivity of the sample (in regard to absorbed optical power) is $7.8\cdot 10^8$ V/W, NEP = $4\cdot 10^{-17}\textrm{W Hz}^{-1/2}$ and the photon noise is $7\cdot 10^{-18}\textrm{W Hz}^{-1/2}$. Taking into account higher working temperature of our sample, we can conclude that our result looks better than previously reported so far.

Finally we have demonstrated that to approach the photon limited mode of operation one should increase the optical power load up to 3-5 pW.

We wish to thank Ernst Otto for help in fabrication of the samples.

The work is supported by Russian Ministry of Science (Project 3.2054.2014/K), RFBR (grant 15-47-02552) and in the framework of Increase Competitiveness Program of Lobachevsky NNSU under contract no. 02.B.49.21.0003.

\end{document}